\documentstyle[preprint,aps]{revtex}

\begin{document}

\draft \title{Zeros of the partition function for a continuum system at
first
order transitions}

\author{ Koo-Chul Lee }

\address{Department of Physics and the Center for Theoretical Physics  \\
         Seoul National University \\ Seoul, 151-742, Korea} \date{\today}
\maketitle

\begin{abstract}

We extend the circle theorem on the zeros of the partition function to a
continuum system.  We also calculate the exact zeros of the partition function
for a
finite system where the probability distribution for the order parameter
is
given by two asymmetric Gaussian peaks.  For the temperature driven first order
transition  in the thermodynamic limit, the
locus and the angular
density of zeros are given by $r = e^{(\Delta c/2l)\theta^2}$ and  $2\pi
g(\theta)=l(1+\frac{3}{2}(\Delta c/l)^2\theta^2)$ respectively in the
complex $z(\equiv re^{i\theta})$-plane where $l$ is the reduced
latent heat, $\Delta c$ is the discontinuity in the reduced specific heat and
$z= \exp(1-T_c/T)$.
\end{abstract}


\narrowtext
One of the fascinating subjects of equilibrium statistical mechanics is to
understand how an analytic partition function acquires a singularity when the
system undergoes a phase transition. For the last three decades main focus has
been  on the second order
transition.   Only recently a renewed interest in the
first order phase transition began to emerge\cite{herr1}.

Since Yang and Lee \cite{leeyang52} first published their celebrated papers on
the theory of phase transitions and the circle theorem on the zeros of the
partition function, there have been many attempts to generalize the
theorem\cite{griffiths73}.  Fisher\cite{fisher65} initiated the study of zeros
of the partition function in the complex temperature plane, and
Jones\cite{jones66} proposed a scenario for the first order transition for a
continuum system.  However very little is known about the distribution of
zeros for the continuum case. This is because  the partition
function for the continuum system is not a polynomial in general and the
original proof of the circle theorem relied heavily on particular
properties of the coefficients of a polynomial.  Recently we have been able to
prove the theorem in a quite different approach\cite{lee94} and this approach
allows us to extend the theorem to the continuum case.

We found that the circle theorem follows from a certain mathematical
relation  which  exists between a probability density
function and the zeros of its characteristic function. In this
paper we first prove three theorems and a corollary without referring to the
partition function.  When these theorems and the corollary are translated in
physical terms we find that,  (1) the zeros of the partition function can be
expressed  in terms of the discontinuities in the derivatives of the free
energy
across the phase boundary if there is a non-vanishing discontinuity
in the first order derivative; (2) there is no zeros in the single phase region
where the probability distribution is given by a single Gaussian
peak; (3) the zeros of the partition function are calculable exactly
at the two phase coexistence point  where the probability distribution is given
by two asymmetric Gaussian peaks; and (4) the zeros lie on the unit circle if
the transition is symmetric.

Furthermore we find  the finite size scaling  very much similar to that
of the discrete system\cite{lee94}.  Therefore this result can again be used,
for a continuum
system, (1) to resolve the recent controversy over equal weight vs.  equal
height of the probability distribution
functions\onlinecite{bind1,bind2,borg90}; and (2) distinguish the first order
transition from the second\onlinecite{lee93,lee94}, just as we have done for
the discrete system in ref.\cite{lee94}.

Consider a probability density function $f(x)$ of a random variable $X$ of
continuous type, which satisfy i) $f(x) \geq 0$ and ii)
$\int_{-\infty}^{\infty}f(x)dx=1$.  The characteristic function\cite{bhat} of
a random variable $X$ is defined by
\begin{equation} \phi(\omega)=
\int_{-\infty}^{\infty}e^{i\omega x}f(x)dx \end{equation} where $i$ is the
imaginary unit.  The logarithm of $\phi(\omega)$ is known as the second
characteristic function or cumulant generating function and denoted by
$\psi(\omega)$.  That is, \begin{equation} \psi(\omega)= \ln(\phi(\omega)) =
\sum_{s=1}^{\infty}\gamma_s\frac{(i\omega)^{s}}{s!}.  \label{eq:cumseries}
\end{equation} The expansion coefficients, $\gamma_s$ are known $s$-th
cumulants
or semi-invariants and are calculable by the formula, $\gamma_s =
\partial^s\psi(\omega)/\partial (i\omega)^s\big|_{i\omega=0}.$

If $\phi(\omega)$ can be analytically continued into the complex
$\omega$-plane,
${\cal M}(t) = \phi(-it)$ is a real function called the moment generating
function.  Since the characteristic function always exists and its
properties are well known\cite{bhat}, we will consider  zeros of
the characteristic function.

Let $S_*$ be a subset of the sample space of a random variable X.  The
conditional distribution of X within the sample space, $S_*$, is called the
{\it
truncated distribution} of $X$, whose probability density function $f_*(x)$ is
given by
\begin{equation} f_*(x) = \frac{f(x)}{\int_{S_*}f(x)dx},
\end{equation}
where $x\in S_*$.  Suppose the characteristic function of the distribution
which can be analytically continued into the complex plane.  We will
consider zeros of the characteristic function in the finite region only.  If
the density function is not a delta function (in that case there are no zeros
in
the finite region) we can always divide the density function into two truncated
density functions, taking a point $x_*$ some point in the middle of the
distribution.

That is,
\begin{equation}
f(x) = c(f_1(x)+af_2(x)),
\end{equation}
where
\[  f_1(x) = \left\{\begin{array}{r@{\quad:\quad}l}
\frac{f(x)}{\int_{-\infty}^{x_*}f(x)dx} &  x<x_* \\
         0 & x\geq x_*~,\end{array} \right. \]

\[  f_2(x) = \left\{\begin{array}{r@{\quad:\quad}l}

 0 & x<x_* \\
 \frac{f(x)}{\int_{x_*}^{\infty}f(x)dx} &  x\geq x_*~,  \end{array} \right. \]

$c = \int_{-\infty}^{x_*}f(x)dx,$ and
$a = \int_{x_*}^{\infty}f(x)dx/\int_{-\infty}^{x_*}f(x)dx.$

Consider the zeros of characteristic function
\begin{equation}
\phi(\omega) = \phi_1(\omega)+a\phi_2(\omega). \label{eq:chsum}
\end{equation}
If  $\psi_1(\omega)$ and $\psi_2(\omega)$, the cumulant generating functions,
of
$\phi_1(\omega)$ and $a\phi_2(\omega)$  exist  except at isolated zeros, then
we
can
write eq.(\ref{eq:chsum}) in terms of these cumulant generating functions as

\begin{equation}
\phi(\omega) = 2e^{\bar{\psi}}\cosh(\tilde{\psi}(\omega)).\label{eq:chfactor}
\end{equation}
In the above
$\bar{\psi}(\omega) = (\psi_1(\omega) + \psi_2(\omega))/2$
and $\tilde{\psi}(\omega) = (\psi_2(\omega) - \psi_1(\omega))/2$.

It should be noted that, because of the factor $a$ in  $\psi_2(\omega)$,
$\psi_2(0)= \ln(a)$. It should further be noted that the
zeros of $\phi(\omega)$, in eq.(\ref{eq:chsum}) are zeros of
$\cosh(\tilde{\psi}(\omega))$ only.  This is because any zeros of
$e^{\bar{\psi}}$ cancel
 the
poles of $\cosh(\tilde{\psi}(\omega))$. This in turn can be understood because
eq.(\ref{eq:chfactor}) is nothing but $2(a\phi_1\phi_2)^{\frac{1}{2}}[\{\phi_1
+a\phi_2\}/2(a\phi_1\phi_2)^{\frac{1}{2}}]$.

Therefore the zeros of $\phi(\omega)$ may be obtained by solving
\begin{equation}
\tilde{\psi}(\omega)= \pm i(1/2+k)\pi\equiv iI_k,  \label{eq:ik}
\end{equation}
where $k=0,1,2,\cdots$.

Now using the cumulant expansion (\ref{eq:cumseries}) we can write
$\tilde{\psi}(\omega)$ as

\begin{equation}
   \tilde{\psi}(\omega)=
\sum_{s=1}^{\infty}\tilde{\gamma}_s\frac{(i\omega)^{s}}{s!}.
\label{eq:cumseriest}
  \end{equation}
where   $\tilde{\gamma}_s = (\gamma_s^{(2)} -\gamma_s^{(1)})/2$.

If
$\tilde{\gamma}_1
=\left[\frac{d}{d(i\omega)}\tilde{\psi}(\omega)\right]_{i\omega=0}\neq 0$,
the inverse function $i\omega=\tilde{\psi}^{-1}(\tilde{\psi})$ exists near the
origin, and the Taylor series about the origin gives, using the Lagrange's
formula\cite{ckp66},
\begin{equation}
    i\omega=
\sum_{s=1}^{\infty}b_s\frac{\tilde{\psi}^s}{s!},
\label{eq:psiseries}
  \end{equation}
where
\begin{equation}
    b_s=
\frac{1}{s!}\frac{d^{s-1}}{d(i\omega)^{s-1}}(\frac{i\omega}{\tilde{\psi}(\omega
)
})^s\Big|_{i\omega=0}.
\label{eq:bcoefficient}
\end{equation}
Notice that (\ref{eq:psiseries}) is the inverted series of
(\ref{eq:cumseriest}). It should also be noticed
that,
since $\tilde{\gamma}_s$'s are real, $b_s$'s are also real. This leads to
theorem I.\\

Theorem I. The zeros of the characteristic function $\phi(\omega)$ can be
expressed by
\begin{equation}
i\omega_k=\sum_{s=1}^{\infty}b_s\frac{(iI_k)^s}{s!},
\label{eq:zeros}
\end{equation}
where $b_s$'s $I_k$ are given by (\ref{eq:bcoefficient}) and  (\ref{eq:ik}).
\\

Although the dividing point $x_*$ is arbitrary and $b_s$'s
should not depend on $x_*$, $\tilde{\gamma}_s$'s have little meaning unless the
original probability density function has two separate distributions over two
distinct regions and $x_*$ is taken at some point between the two regions.

For a symmetric distribution  where $f(x^*+x) =
f(x^*-x)$, it is convenient to consider a random variable defined by
$Y=X+x^*$. Then the density function becomes symmetric about the origin,
$\hat{f}(-y)=\hat{f}(y)$ and  the zeros of $\phi_y(\omega)$ are the same as
those of  $\phi_x(\omega)$ since
$\phi_y(\omega)=e^{ix^*\omega}\phi_x(\omega)$.

It can be easily shown that the characteristic function is symmetric  about the
origin iff  the density function is symmetric.
Let us take for a symmetric density function $f(y)$ about the origin and the
corresponding characteristic functions of truncated distribution functions,
$\phi_1(\omega) =
2\int_{-\infty}^{0}e^{i\omega y}f(y)dy$
and $\phi_2(\omega) = 2\int_{0}^{\infty}e^{i\omega y}f(y)dy$.
Since $\phi_1(-\omega) = \phi_2(\omega)$,
we have $\tilde{\gamma}_s = \gamma_s^{(2)} -\gamma_s^{(1)}= 0$,
for even $s$, and $\tilde{\gamma}_s = \gamma_s^{(2)} =
-\gamma_s^{(1)}$, for odd $s$.

Therefore
\begin{equation}
  \tilde{\psi}(\omega) =
\sum_{s=0}^{\infty}\tilde{\gamma}_{2s+1}\frac{(i\omega)^{2s+1}}{(2s+1)!}=
i\sum_{s=0}^{\infty}(-1)^s\tilde{\gamma}_{2s+1}\frac{\omega^{2s+1}}{(2s+1)!}.
  \end{equation}
By the inverse series given by eq.(\ref{eq:psiseries}), we have
\begin{equation}
   \omega=
\sum_{s=0}^{\infty}(-1)^sb_{2s+1}\frac{(\tilde{\psi}/i)^{2s+1}}{(2s+1)!},
  \end{equation}
Substituting the solution for zeros (\ref{eq:ik}) we have
\begin{equation}
    \omega_k=
\sum_{s=0}^{\infty}(-1)^sb_{2s+1}\frac{I_k^{2s+1}}{(2s+1)!},
\label{eq:tk}
  \end{equation}
provided that series converges.
  Since $b_s$'s are real $\omega_k$'s are real. This leads to
the following  corollary.\\

Corollary. ~  Zeros of characteristic function of a symmetric distribution
function  lie on  the real axis, provided that the series in (\ref{eq:zeros})
converges.\\

  This means that in the complex
$z=e^{-i\omega}$-plane, the Mellin
transformation
$P(z)$, defined by $P(z)=\phi (-i\ln(z))$\cite{bhat} has zeros only on the unit
circle.

For some probability distributions we show next that it is possible to
calculate zeros of the characteristic function explicitly.

If the density function of the Gaussian distribution is
$f(x)=\exp(-\frac{(x-\mu)^2}{2\sigma^2})/\sqrt{2\pi}\sigma$, the cumulant
generating function is given by $\psi(\omega)=
\mu(i\omega)+\frac{1}{2}\sigma^2(i\omega)^2$\cite{bhat}.
Since the exponential function of
an arbitrary entire function can not have  zeros in the finite
region \cite{ckp66} and  $\psi(\omega)$ in the above is an
entire function, $\phi(\omega)=e^{\psi(\omega)}$ can not have zeros in the
finite region. This leads to theorem II.\\

Theorem. II. ~ The characteristic function of the Gaussian distribution has no
zeros
in the finite region.\\

On the other hand if the density function of a double Gaussian
peak is given by
\begin{equation}
f(x)=e^{-\frac{(x-\mu_1)^2}{2\sigma_1^2}}/\sqrt{2\pi}\sigma_1
+ae^{-\frac{(x-\mu_2)^2}{2\sigma_2^2}}/\sqrt{2\pi}\sigma_2,
\label{eq:double}
\end{equation}
the cumulant generating functions for the two  peaks are given by
$\psi_1(\omega)=\mu_1(i\omega)+\frac{1}{2}\sigma_1^2(i\omega)^2$, and
$\psi_2(\omega)=\ln(a)+\mu_2(i\omega)+\frac{1}{2}\sigma_2^2(i\omega)^2$
respectively.
Therefore $\tilde{\psi}(\omega)=
\ln(a)/2+m(i\omega)+\frac{1}{2}\tilde{\sigma}^2(i\omega)^2$, where
$m=(\mu_2-\mu_1)/2$ and $\tilde{\sigma}^2=(\sigma_2^2-\sigma_1^2)/2$.
Solving the equation (\ref{eq:ik}) using the above $\tilde{\psi}(\omega)$,
the zeros of the characteristic function can be obtained explicitly after
performing some algebra. This leads to theorem III.\\

Theorem. III. ~  If the density function of a double Gaussian
peak is given by (\ref{eq:double}), the zeros of the  characteristic function
are
$\omega_k=\lambda_k/|\tilde{\sigma}|
+i(m-|\tilde{\sigma}I_k/\lambda_k|)/\tilde{\sigma}^2$, where $\lambda_k =\pm
[\{(\frac{m^2}{2\tilde{\sigma}^2}-\frac{\ln(a)}{2})^2
+I_k^2\}^{\frac{1}{2}}
-(\frac{m^2}{2\tilde{\sigma}^2}-\frac{\ln(a)}{2})]^{\frac{1}{2}}$.\\

In the asymptotic limit where $a\rightarrow 1$ and
$\tilde{\sigma}\rightarrow 0$, the zeros are given by
\begin{equation}
\omega_k=I_k/m/(1-\epsilon_k\tilde{\sigma}^2/m) + i\epsilon_k
\label{eq:tkg}
\end{equation}
where $\epsilon_k = \ln(a)/2m
-\frac{1}{2}\tilde{\sigma}^2I_k^2/m^3$.  If we put $a = 1$ and $\tilde{\sigma}
=
0$ in the above, then we have $\epsilon_k=0$, which makes  $\omega_k$'s real.
This is
an explicit example of  the corollary of theorem I.

The above mathematical results can be readily applied to the theory of the
phase
transition. As an example consider a canonical partition function $Z(\beta)$
defined by
$Z(\beta)= \int_{-\infty}^{\infty}e^{-\beta E} \Omega(E)dE$, where $\Omega(E)$
is the density of states at $E$, $\beta$ is the inverse temperature
$1/k_bT$ and $k_b$ is the Boltzmann's constant. We can identify this partition
function as a
moment-generating function ${\cal M}(t)$,
\begin{equation}
 {\cal M}(t) \equiv Z(\beta)/Z(\beta_o)=\int_{-\infty}^{\infty}e^{tx}f(x)dx
\end{equation}
where the probability density function is given by
\begin{equation}
f(x)= \Omega(x/\beta_o)e^{-x}
/\int_{-\infty}^{\infty}\Omega(x/\beta_o)e^{-x}dx.
\end{equation}
In the above $t=1-\beta/\beta_o$ , $x=\beta_o E$,
and $\beta_o$ is a reference inverse temperature around which the system
fluctuates.

Since the cumulants of $f(x)$ are related to the derivatives of the free
energy,
theorem I implies that the zeros of the partition function can be expressed in
terms
of the discontinuities in the derivatives of the free energy provided that the
first
order derivative has a non-vanishing discontinuity.
If $f(x)$ is a symmetric function, then  the corollary of theorem I says that
the
zeros of the partition function lie on the unit circle in the complex
$e^t$-plane. Thus we have extended the Lee-Yang unit circle theorem to the
continuum
case.
{}From the general principle of statistical mechanics, $f(x)$ can be
approximated
by the Gaussian distribution.  In the
thermodynamic limit its mean is given by the
internal energy $\beta_oU$, and the variance by the heat capacity as $\sigma^2=
C_v/k_b$.  Here
$\beta_o$ is the inverse temperature of a single phase.
Due to theorem II, the partition function has no zeros.

On the other hand  if the system undergoes a first order transition at
$\beta_o=\beta_c$,
then $f(x)$ is characterized by a double Gaussian peak
\onlinecite{bind1,bind2} separated by the
discontinuity of the internal energy, or the latent heat in the thermodynamic
limit.
Let us further assume that the ratio of the weight of the two peaks is $a$, and
there is also a discontinuity in $C_v$, $\Delta C_v$. Let $N$ be
some integer representing an extensive thermodynamic quantity, say the number
of
particles of the system.  Introducing the reduced internal energy
$u=\beta_cU/N$,  the reduced latent heat
$l = \Delta u \equiv u_2-u_1$ and the reduced specific heat $c= C_v/Nk_b$,
theorem III now says that zeros of the partition function for small values of
$k$ may be written as,
\begin{equation}
\ln(r_k)=\Re (t_k) = -\ln(a)/Nl + \frac{1}{2}(\Delta c/l)(\vartheta_k/l)^2
\end{equation}
\begin{eqnarray}
\theta_k=\Im (t_k)
= (\vartheta_k/l)\{&&1 + \frac{1}{2}(\Delta c/l)\ln(a)/Nl\nonumber\\
&& - \frac{1}{2}(\Delta c/l)^2(\vartheta_k/l)^2\}
\end{eqnarray}
Here we have used  $\tilde{\sigma}^2= \Delta cN/2$, $m=lN/2$ and $\vartheta_k=
(1+2k)\pi /N$ with $k=0,\pm 1,\pm 2,\cdots$.
We see that the dominant finite size correction is the terms dependent on
the asymmetric factor  $a$.

In the thermodynamic limit where $N\rightarrow \infty$,  $a$
dependent terms  vanish, and the equation for the locus of zeros becomes
\begin{equation}
r = e^{(\Delta c/2l)\theta^2},
\label{eq:locus}
\end{equation}
where $r_k$  and $\theta_k$ are replaced by  the continuous variables
$r$ and  $\theta$. In Fig.~1, we plot $u$ as a function $T/T_c$, and zeros in
the complex $t-$ and $e^t-$planes. $l=1.0$ and $\Delta c=\pm 0.2, 0$ are used
in
all three figures and the exact zeros given by theorem III are calculated using
$N=20$ and $a=1$.

The angular density of zeros defined by
$Ng(\theta)=1/(\theta_{k+1}-\theta_k)$ can be written for
small values of $k$, as
\begin{equation}
2\pi g(\theta)=l\{1+\frac{3}{2}(\Delta c/l)^2\theta^2\}.
\end{equation}
If the transition is symmetric ($\Delta c =
0$), the locus of zeros becomes the unit circle with the uniform density $2\pi
g(\theta)=l$.
One should note that the equation for the locus of zeros is valid only near the
real axis. This is because terms beyond the Gaussian approximation become
important as the argument $\theta$ grows, {\it i,e} for large
values of $k$,  as we have shown with an example in  ref.\cite{lee94}.
Finially
it should be remarked that the number of zeros in this example is infinite in
the  complex $t$-plane (Fig.~1(b)). However
only a finite number of zeros closes the circle in the complex  $z= e^{t}$ for
a
finite system if we consider only the first Riemann sheet. In fact for the
symmetric case where $\Delta c = 0$, there are exactly $N$ zeros distributed
uniformly on the unit circle if we scale the energy of the system by $l$, so
that $l=1.0$.

In conclusion we have shown that the scenario for the  first order phase
transition put forward  by Yang and Lee\cite{leeyang52} is valid in a continuum
system. Becuase we have shown a formal relation between the
discontinuities in the derivatives of the free energy and the distribution of
zeros of the partition function,  it can now be applied to any type of first
order phase transition. It can also be extended to a multi-phase
coexistence point. In this case,  one only needs to consider
the multi-dimensional
complex space as we have done in ref.\cite{lee94}. The existence of the
formal relation presented in this paper was suspected by the original
proponents of the theorem Lee and Yang themselves. In their 1952
paper\cite{leeyang52}, they expressed their sentiment in the concluding remark
by saying, ``\ldots distribution (of zeros) should exhibit such simple
regularities \ldots . One can not escape the feeling that there is a very
simple basis  underlying the theorem, with much wider application, which still
has to be discovered.''  We believe we have discovered this simple basis.

This work was supported in part by the Ministry of Education, Republic of
Korea
through a grant to the Research Institute for Basic Sciences, Seoul National
University, in part by the Korea Science Foundation through Research Grant to
the Center for Theoretical Physics, Seoul National University and in part  by
S.N.U.  Daewoo Research Fund.

\begin{figure}
\caption
{(a) $u$ as a function of
$T/T_c=1/(1-t)$. $u_1=0.7$ and  $c_1=1.0$ are taken arbitrarily. (b) Zeros  in
the complex $t$-plane
given by theorem III.   For $\Delta c > 0$, the line of zeros arches toward the
positive real axis. Only zeros
between the two dashed lines appear in the first Riemann sheet in the complex
$e^t$-plane.
(c) Zeros in the complex $e^t$-plane. The solid lines are the loci of the zeros
in the thermodynamic limit given by (21).  For $\Delta c > 0$, the locus lies
outside the unit circle which corresponds to the symmetric case, $\Delta c =
0$.
 }
\label{fig1}
\end{figure}

\end{document}